\documentclass[10pt,conference]{IEEEtran}
\IEEEoverridecommandlockouts
\usepackage{cite}
\usepackage{amsmath,amssymb,amsfonts,bm}
\usepackage{algorithmic}
\usepackage{xcolor}
\usepackage{graphicx}
\usepackage{multirow}
\usepackage{comment}

\usepackage[linesnumbered,ruled,noend]{algorithm2e}

\SetAlgoCaptionSeparator{:}
\SetNlSty{texttt}{}{:}
\usepackage[T1]{fontenc}
\usepackage[font=small,labelfont=bf,tableposition=top]{caption}

\DeclareCaptionLabelFormat{andtable}{#1~#2  \&  \tablename~\thetable}

\SetAlgoCaptionSeparator{:}
\SetNlSty{texttt}{}{:}

\newcommand{\flowminer}{\emph{FlowMiner}\xspace}
\newcommand{\perracotta}{\emph{Perracotta}\xspace}
\newcommand{\texada}{\emph{Texada}\xspace}
\newcommand{\tlmine}{\emph{TLMine}\xspace}
\newcommand{\prefixspan}{\emph{PrefixSpan}\xspace}
\newcommand{\ylstm}{\emph{YLSTM}\xspace}
\newcommand{\gt}{\emph{GT}\xspace}

\def\BibTeX{{\rm B\kern-.05em{\sc i\kern-.025em b}\kern-.08em
    T\kern-.1667em\lower.7ex\hbox{E}\kern-.125emX}}
\begin{document}

\title{ Tools and Algorithms for SoC Communication Traces\\
}

\author{
\IEEEauthorblockN{Md Rubel Ahmed}
\IEEEauthorblockA{
U. of South Florida, FL, USA\\
mdrubelahmed@usf.edu}
\and
\IEEEauthorblockN{Hao Zheng}
\IEEEauthorblockA{
U. of South Florida, FL, USA \\
haozheng@usf.edu}
 }

\maketitle

\begin{abstract}

In this paper we study seven well-known trace analysis techniques both from the hardware and software domain, and discuss their performance on communication centric system-on-chip (SoC) traces. SoC traces are usually huge in size and concurrent in nature, therefore mining SoC traces poses additional challenges. We provide a hands on discussion of the selected tools/algorithms in terms of the input, output and analysis method they employ. Hardware traces also varies in nature when observed in different level, this work can help developers/academicians to pick up the right techniques for their works. We take advantage of a synthetic trace generator to find the interestingness of the mined outcomes for each tool as well as we work with a realistic GEM5 setup to find the performance of these tools on more realistic SoC traces. Comprehensive analysis of the tools performance and a benchmark trace dataset are also presented.





\end{abstract}

\begin{IEEEkeywords}
SoC validation, trace analysis, specification mining, SoC execution model, system-on-chip
\end{IEEEkeywords}


\section{Introduction}
\label{sec:motivation}

Transaction level SoC design is becoming trivial to keep up with the pace of ever increasing consumer demand. This is accelerated the communication centric design verification as rtl level simulation verification does not scale up with the complexity. 
Trace analysis or logging is a widely used debug method for both complex and large hardware and software domains. Tracing has enabled many large systems to grow and function correctly to these days.
Today's computing systems are so complex that even the SoC in our wrist watch can not be debugged manually. 
Therefore the attempts to automate the trace analysis has gained momentum decades ago. Many tools and algorithms have been introduced to tackle the problem from different angles. A methodical comparison of such tools solely targeting SoC communication traces is necessary.

Modern trivial high performing SoC houses a variety of intellectual property (IP) blocks, such as memories, network interfaces, direct memory controllers etc. Therefore communication centric debug has been proven more practical than traditional computation centric debug methods~\cite{transaction_based_07}. High volume of parallel transaction capacity is a trivial demand from a modern interconnect that manages the realization of various on-chip protocols. Therefore, the IP communication data captured from the different interconnect is very complex in nature but reveals many opportunities for test and debug.

\begin{figure}[tb]
\begin{center}
\includegraphics[width=.34\textwidth]{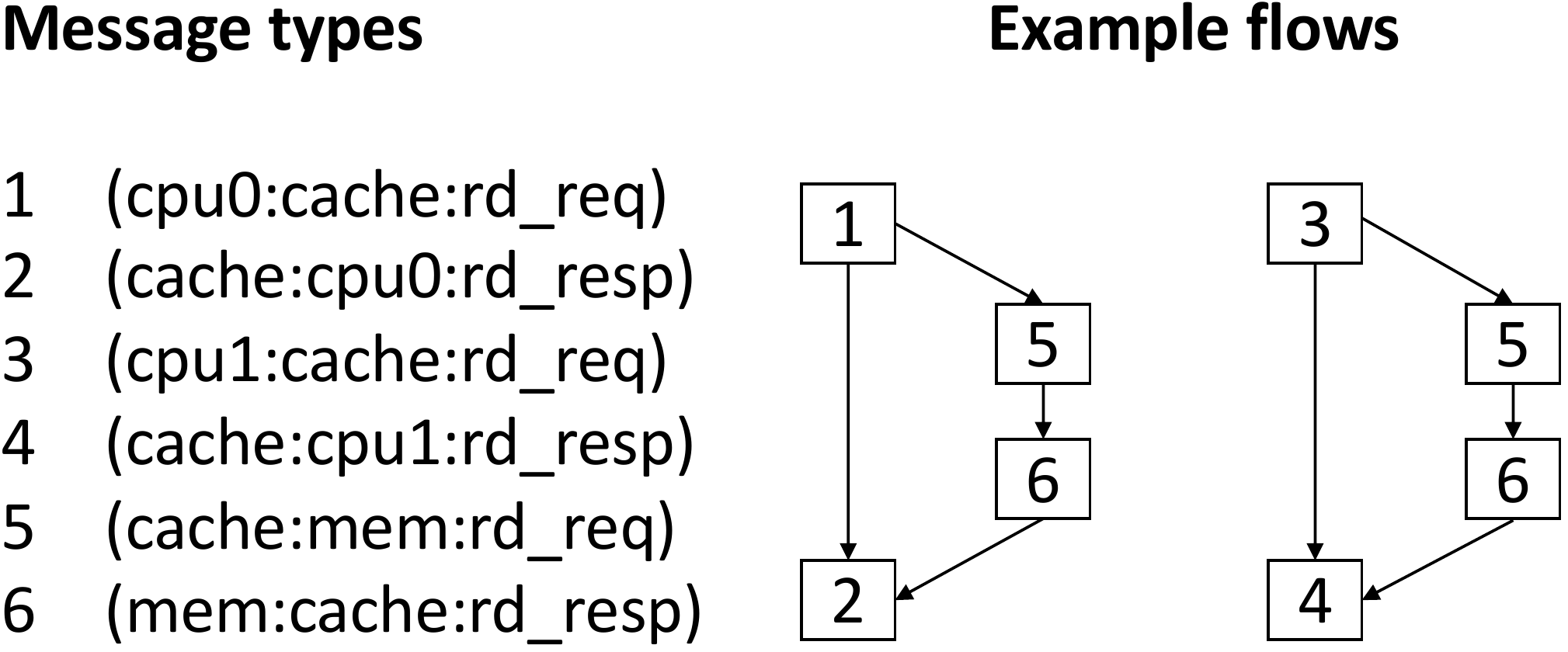}
\caption{A CPU downstream write flow~\cite{ahmed2020mining}.}
\label{fig:flow-ex}
\end{center}
\vspace*{-6mm}
\end{figure}

An example communication interaction between IPs is presented in Fig.~\ref{fig:flow-ex}. Parallel execution of this CPU dowonstream write activity can generate a trace in eq.~\ref{ex:trace} where the numbers in the trace are the message indices referred as message or event\footnote{Message and event will mean similar throughout the paper}. Trace mining algorithm targeting reconstruction of the flow specification in the form of the sequential patterns or \emph{Ground Truths \textit{GT}}\footnote{The exact order of message execution is referred as \textit{GT}} patterns such as \{1,2\}, \{1, 5, 6, 2\}, \{3, 4\}, \{3, 5, 6, 4\} can automate the manual and unscalable job of writing and updating flow specifications in real electronic design automation.

An example trace from executing the flows in Fig.~\ref{fig:flow-ex} is 
\begin{equation}
\label{ex:trace}
\left(\{1, 3\}, 1, 2, 5,  1, 5, 6, 2, 4, 6, 2\right)
\end{equation}
where the numbers in the curly braces indicates that the transaction {\tt cpu0:cache:rd\_req} and {\tt cpu1:cache:rd\_req} were captured in the same timestamp.

SoC IP communication traces are diverse, concurrent and large volume in nature. We work with realistic SoC communication traces for seven tools and find their strong and weak points for such traces. The purpose of this work is to answer the question that how different tools performs for finding meaningful communication patterns.
We present our work to answer the following questions as follow:
\begin{enumerate}
    \item What tool deals best with SoC communication traces?
    \item Which tool/s can handle large and concurrent communication traces?
    \item Why some tools performs better (based on a few performance matrics)?
\end{enumerate}

\noindent To the best of our knowledge, this is the first paper that presents a comparative study trace analysis tools for SoC transaction level traces. The \textbf{contribution}s are as follow:

\begin{enumerate}
    
    \item A set of traces that can work as a communication trace benchmark datasets
    \item A comprehensive evaluation of trace analysis tools and algorithms on the benchmark
\end{enumerate}

\noindent Following sections discuss classification, short overview, benchmark generation, strengths and weakness of different tools sequentially.
\section{Classification of Trace Miners}
\label{sec:classification}

We aim towards tools and techniques that mines patterns or extracts models from the traces. This list includes many tools ranging both from the software and hardware domains. As the necessity of this test and debug increase, more works in this line are expected to roll out in near future. However, we study several well known tools and techniques in this paper. The classification of the existing tools broadly depends on their underlying principle and methods. We can loosely separate them in following classes but a fine margin is often a very hard one to draw between them. We list them separately only for the reason to select representative work from each category.

\noindent {\bf Automata Based Miners } 
Automaton based approaches constructs FSMs(Finite State Machine) that can interpret the event causality observed in the traces. The states are often the unique events found in the trace and the transition are the relations between the states which are solved using SAT solvers. Prominent work in this category includes \emph{Trace2Model}~\cite{natasa2020}, model synthesis work~\cite{zheng2021model}, Mining Specifications~\cite{mining_specs_ammons2002} that propose tools and techniques to construct automatons from different types of traces. Model Synthesis from SoC IP communication traces described in~\cite{zheng2021model} builds causality graphs using structural features of the messages.A SAT is used to solve the edges depending on the support information calculated from the trace. We consider \emph{Trace2Model}  from this category of tools to be studied in this paper.

\noindent {\bf LTL Miners} 
A model building approach \textit{Synoptic} \cite{Beschastnikh:2011:LEI:2025113.2025151} first mines invariants from logs of sequential execution traces where concurrency is recorded in partial order. It then generates a FSM that satisfies the mined invariants. It builds execution models to enhance programmer comprehension of the traces. Another tool called \texada~\cite{texada_main} works with user specified templates in the form of Linear Temporal Logic (LTL)~\cite{book:ltl_book} and produces instances of that formula using some interestingness measures. \texada is considered as the representative tool from this category.   

\noindent {\bf Rule based Pattern miners} 
Some works from this category are~\cite{wencho_dac2010}, \perracotta~\cite{perracotta2006}, \flowminer~\cite{ahmed2020mining}. These works scans the traces to form rules that can interpret the trace. Work \flowminer can find patterns from complex SoC trace and tool \perracotta is a benchmark tool for many other tools. So both of them are considered to be studied in this paper.

\noindent {\bf Assertion based Miners}
Assertions are interesting and useful in SoC debug. With the advent of SystemC, assertions from the SoC traces have been a prime tool for design verification. Assertions are often extracted in the form of sequential patterns~\cite{auto_asser_aspdac2010},~\cite{Liu:2013}. We consider~\cite{Liu:2013} as a representative tool which we refer as \tlmine.

\noindent {\bf General Sequential Pattern miners}
This line of works collectively search for algorithms to discover patterns in sequential data~\cite{FournierViger2017ASO}. This field becomes a very active research area after Agrawal and Srikant~\cite{agrawal1994} proposed their works on frequent itemset mining named coined the term associate rule mining. Mining sequential patterns from sequence database is a broad research area. There exists a handful of well algorithms designed for mining sequential pattern mining. Among many wellknown algorithms, \prefixspan~\cite{prefixSpan2004} is considered.

\noindent {\bf Machine Learning Methods} Machine Learning algorithms have already been contributing solving exciting pattern mining works. The core idea behind applying machine learning based model is that the traces we get posses strong temporal relations among the events. Therefore Recurrent Neural Networks (RNN) have become a prime tools to find the probabilistic relations among the events, finally forming patterns. An example work presented in~\cite{bayspec_dac2019} that utilizes Baysian Inference to interpret the execution model using LTL formula. ~\cite{yuting_LSTM}, ~\cite{deep_spec_2017} are example work in this line of work. We will take~\cite{yuting_LSTM} for our consideration.
\section{Selected Trace Miners}
\label{sec:existing_miners}

To the best of our knowledge, there are not many tools available that directly target SoC communication traces and looks for interesting patterns. 
Therefore, we primarily focus on the generality of each tool that deals with trace analysis. 
We go over the short description of the tools listed below and summarize the tools in Table~\ref{tab:tool_feature}.
\subsection{PrefixSpan}
\noindent This is a sequential pattern mining algorithm that falls into the class of pattern-growth algorithms which employ depth-first-search~\cite{FournierViger2017ASO} technique to explore the candidate pattern search space. It takes minimum support threshold to identify frequent patterns of single item in the sequence database. Then it generates longer patterns appending items to the prefix or suffix of the pattern under consideration. This algorithm comes with huge cost in memory but avoid searching for patterns that do not exist in the database. We obtain the implementation of this algorithm that is ported with the open-source data mining library called SPMF~\cite{spmf2014}.
For the trace in eq.~\ref{ex:trace} \prefixspan could find 2463 patterns which also includes four \gt patterns \{1,2\}, \{1, 5, 6, 2\}, \{3, 4\}, \{3, 5, 6, 4\} which is interesting.

\subsection{Trace2Model}
\begin{figure}
    \centering
    \includegraphics[width=3.4in]{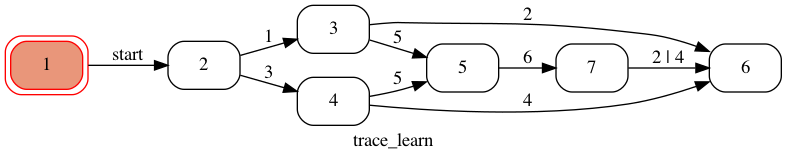}
    \caption{FSA model produced using the tool \emph{Trace2Model}}
    \label{fig:trace2model-output}
\end{figure}

\noindent It works on system execution traces and there is learning model that synthesizes the traces into a very concise and abstract model. The model is an automaton with a set of transitions in which each set element has current state, transition predicate and the next state. It solves the edge transitions as a boolean (SAT) solving problem. We utilize the incremental searching version of the tool. It starts with N states and then looks for (N+1)$^{th}$ state to be satisfied by a C Bounded Model Checker (CMBC). The model helps to understand the system behavior under a particular program execution. It utilizes sliding window technique as an optimization method for scalability. Other than the window size \textit{w}, there is another hyperparameter \textit{l} that controls the degree of generalisation of the learned automaton. 
It shows significant improvements over state merge algorithms for generating behavioral models. The model in Fig.~\ref{fig:trace2model-output} is resulted from the trace in eq.~\ref{ex:trace}. It can capture the transitions successfully. 

\subsection{TLMine}
\noindent \tlmine is an episode mining framework that mines assertions in the form of frequent episodes from transaction level model (TLM) simulation traces. The algorithm works for longer episodes in an incremental manner. It describes an abstraction of communication actions as events. An episode is an ordered sequence of events. It looks for episodes of multiple elements for a given support\footnote{Please refer to the original paper for definition.} and confidence thresholds.  
An episode is called frequent if it appears in certain number of time windows which it call the support value of an episode. It introduces the confidence of an episode which is the ration between the supports values of an episode and its prefix. As they mine assertions, 100\% confidence value is employed for an episode to be considered. It shows benefit of applying sliding window technique over sequential pattern mining in terms of run time and number of episodes mined. For the trace in eq.~\ref{ex:trace}, this tool could find 30 episodes. Among them \{1, 2\} is present as the only \gt pattern.

\subsection{FlowMiner}
\noindent \flowminer can find meaningful patterns from the IP communication traces of an SoC executions. These patterns are supposed to be utilized as flow specifications for the IP communications. The speciality of this tool is that it mines patterns from highly concurrent and interleaved traces. The input of this tool is a set of execution traces over messages observed in various communication interfaces in an SoC design. It utilizes the well-known data mining support confidence framework to find binary rules which are also invariant over the traces. Further these mined binary rules are chained to longer patterns through some inference techniques while patterns are could also be treated as invariants. The associated paper also shows some optimization techniques to reduce the mining complexity as well as improve the mined patterns quality. From the example trace, it could find all the four \gt patterns along with 7 additional patterns.

\subsection{Perracotta}
\noindent This work mines temporal properties in the form of API rules from the program execution traces. It mines properties based on some user defined templates. There is an inference engine that works on the input traces incorporating a set of property templates. For each property template, there is an FSM that is used to scan the trace to find the instance of that property. This FSM plays a vital role for making this tool scalable to large set of trace.

The inferred properties go for further post processing to finally be reported to the user. There are eight different property templates which are abstracts of a set of concrete properties. At first \perracotta targets mining temporal properties of two events or values. Among the 8 property types, we find alternating properties are most interesting for the SoC communication pattern mining context as patterns are also invariants. Therefore, we further restrict our discussion on this tool about the alternating properties only. The paper also describes a technique to produce \textit{Alternating chains} that is the composition of alternating properties to deduce complex FSMs out of smaller properties. This saves huge computation inherent to finding longer patterns from the sequential traces. 

The inference engine uses a metric called satisfaction rate a threshold to rank alternating properties. A trace is partitioned into small sub-traces and then a conformity check is done in these sub-traces that provides the satisfaction rate. 14 rules are found from the example trace in different template patterns. Among them \{3, 4\} exists as the only \gt pattern in the form of alternating patterns.

\begin{table*}[!t]
\renewcommand{\arraystretch}{1.3}
\caption{Summary of the tools under consideration}
\label{tab:tool_feature}
\centering
\begin{tabular}{|l||l|l|l|l|}
\hline
{\bf Tool}        &  {\bf Application }    & {\bf Method}    & {\bf Input}            & {\bf Output}     \\ \hline \hline

\flowminer  & \begin{tabular}[c]{@{}l@{}} Message flow mining, design\\automation, SoC validation,\\Work flow mining, Specification\\ inference  \end{tabular} & \begin{tabular}[c]{@{}l@{}} Association rule mining,\\inference for chaining, use of\\support-confidence framework \end{tabular} & \begin{tabular}[c]{@{}l@{}} SoC concurrent communication\\traces, Support confidence\\threshold \end{tabular} & \begin{tabular}[c]{@{}l@{}}Flow specifications\\ in the form of\\ sequential patterns\end{tabular} \\ \hline

\perracotta  & \begin{tabular}[c]{@{}l@{}} API rule mining, Mining\\ program behavior, Program\\correctness checking  \end{tabular} & \begin{tabular}[c]{@{}l@{}} Uses property templates and\\looks for template instances using\\FSMs. Uses chaining to obtain\\complex rules \end{tabular}    & \begin{tabular}[c]{@{}l@{}} Set of traces, Concurrency\\not considered satisfaction\\rate \end{tabular} &\begin{tabular}[c]{@{}l@{}} property instances, \\ patterns   \end{tabular}   \\ \hline

\prefixspan & \begin{tabular}[c]{@{}l@{}} Sequential data analysis, click\\stream, CyberSecurity \end{tabular}& \begin{tabular}[c]{@{}l@{}} recursively grows frequent\\patterns using depth first search,\\uses projected database \end{tabular} & \begin{tabular}[c]{@{}l@{}} Sequential data, takes multiple\\traces support threshold \end{tabular} & \begin{tabular}[c]{@{}l@{}} Sequential patterns\\for that threshold \end{tabular}\\ \hline

\tlmine  & \begin{tabular}[c]{@{}l@{}} Digital designs, assertion based\\ verification, transaction level\\models \end{tabular}& \begin{tabular}[c]{@{}l@{}} Mines episodes incrementally\\for a given support, confidence\\threshold, uses maximum\\lifetime of a transaction for\\windowing \end{tabular}  & \begin{tabular}[c]{@{}l@{}} TLM level traces, parallel\\executions are interleaved in\\the traces, support, confidence\\threshold, maximum life\\time of a transaction  \end{tabular} &  \begin{tabular}[c]{@{}l@{}} assertions in\\ the form of\\ frequent episodes\\ of different length \end{tabular} \\ \hline

\texada & \begin{tabular}[c]{@{}l@{}}Program behavior analysis,\\model correction checking,\\mining relation between\\ the events in the program\\execution logs\end{tabular}  & \begin{tabular}[c]{@{}l@{}}Uses template, looks for\\ template instances, checks for\\counter example over the traces \end{tabular}    &    \begin{tabular}[c]{@{}l@{}} Multiple traces, concurrency\\is expressed as interleaving\\ of events in the traces, LTL\\formula or templates according\\to the user's interest  \end{tabular} & \begin{tabular}[c]{@{}l@{}} LTL formula\\ instances\end{tabular} \\ \hline

\emph{Trace2Model} & \begin{tabular}[c]{@{}l@{}} Model building from the\\program execution traces for\\hardware or software system \end{tabular} & \begin{tabular}[c]{@{}l@{}} builds nfa and then increments \\the states until counter example\\ is found in the trace, finally\\converts the nfa to dfa \end{tabular}  & \begin{tabular}[c]{@{}l@{}} Sequential traces, concurrency\\is not considered, degree of\\generalization \end{tabular} &  \begin{tabular}[c]{@{}l@{}} FSM model that\\conforms to the\\observed traces\end{tabular} \\ \hline

\ylstm & \begin{tabular}[c]{@{}l@{}} SoC validation, SoC system\\ protocol specification mining,\\ electronic design automation  \end{tabular} & \begin{tabular}[c]{@{}l@{}} trains LSTM networks to find the\\sequential dependency among the\\events in a trace, incrementally\\ extracts sequential patterns from\\the LSTM models \end{tabular}  & \begin{tabular}[c]{@{}l@{}} SoC system level transaction\\ traces, probability threshold \end{tabular} &  \begin{tabular}[c]{@{}l@{}} Sequential patterns\\ that can describe the\\ SoC system protocol\\specification\end{tabular} \\ \hline

\end{tabular}
\end{table*}

\subsection{Texada}
\noindent \texada mines program behaviour in the form of LTL formulae from the logs or traces. The input to \texada is an LTL property template and log of traces where sequence of string events are placed in the temporal order they appeared in the system. The tool outputs a set of property instantiations that are true over the traces. It utilizes an efficient representation of the input trace that ensures unnecessary traversal over the trace will be avoided. Property instances are validated over the trace in a linear recursive fashion.
\texada also includes some optimization technique such as state memoization to further reduce the search complexity of the tool. We use the LTL formula {\tt G(x $\xrightarrow{}$ X(F(y)))} is to find all the instances of x, y where {\tt "x is always followed by y"}. For this template pattern, \texada finds 15 rules. Among them (1, 2), (1, 5), (1, 6), (3, 4), (3, 5), (3, 6), (5, 6), (6, 2) are interesting as they conform to the flow instances.

\subsection{YLSTM}
\noindent Work~\cite{yuting_LSTM} (we refer as \ylstm) applies LSTM (Long Short Term Memory based on RNN) to extract sequential patterns from SoC transaction-level traces. LSTM is capable of capturing "long-term" dependencies and has many applications in natural language processing. Mined patterns resembles dependencies between various events. This paper exclusively focuses on the concurrent nature of SoC traces.  At first a set of LSTM networks are trained on the SoC traces for different length set by the user.  This trained networks can predict next event upon given an input sequence. To reduce the run-time, all input sequence of single events are considered by the first LSTM model. The output of this model is then feed to the other models for extension incrementally. Finally patterns are extracted from the trained model using another algorithm. As an LSTM model needs enough training data, the single trace in the example is not considered for this tool.

\begin{table*}[!t]
\renewcommand{\arraystretch}{1.3}
\caption{Tool Input Specifications. {\color{red} table will be removed, using for tracing reference}}
\label{tab:tool_input_specs}
\centering
\begin{tabular}{|l|c|c|c|c|c|c|}
\hline
\multicolumn{1}{|c|}{\textbf{Tool}} & \textbf{Trace format}                                                 & \textbf{Event sep.} & \textbf{Trace sep.} & \textbf{Mul. Traces} & \textbf{Con.Traces} & \textbf{|event\_set|} \\ \hline \hline
\textit{\textbf{FlowMiner}}                & 0 -1 1 2 -1 3 -2                                                      & -1                  & -2                  & yes                  & yes                 & $\geq$1      \\ \hline
\textit{\textbf{Perracotta}}             & \begin{tabular}[c]{@{}c@{}}0\\ 1\\ 2\\ 3\\ - - - -\end{tabular}          & new line            & - - - -                & yes                  & yes                 & 1                     \\ \hline
\textit{\textbf{PrefixSpan}}                & 0 -1 1 2 -1 3 -2                                                      & -1                  & -2                  & yes                  & yes                 & $\geq$1      \\ \hline
\textit{\textbf{TLMine}}                & 0 -1 1 -1 2 -1 3 -2                                                   & -1                  & -2                  & yes                  & yes                 & 1                     \\ \hline
\textit{\textbf{Texada}}            & \begin{tabular}[c]{@{}c@{}}0\\ 1\\ 2\\ 3\\ . .\end{tabular}            & new line            & . .            & yes                  & yes                 & 1                     \\ \hline
\textit{\textbf{Trace2Model}}               & \begin{tabular}[c]{@{}c@{}}start\\ 0\\ 1\\ 2\\ 3\\ start\end{tabular} & new line            & start               & yes                  & no                  & 1                     \\ \hline
\textit{\textbf{LSTM-Yuting}}              & 0 1 2 3                                                               & space               & new line            & yes                  & yes                 & 1                     \\ \hline
\end{tabular}
\end{table*}

\begin{table*}[!t]
\renewcommand{\arraystretch}{1.3}
\caption{Mining Summary}
\label{tab:mining_summary}
\centering
\begin{tabular}{|l||c|c|c|c|c|c|c|c|c|c|c|c|}
\hline
\textbf{Trace}          & \multicolumn{2}{c|}{\textit{\textbf{SNI}}}                   & \multicolumn{2}{c|}{\textit{\textbf{SI}}} & \multicolumn{2}{c|}{\textit{\textbf{MI}}} & \multicolumn{2}{c|}{\textit{\textbf{FS}}} & \multicolumn{2}{c|}{\textit{\textbf{Snoop}}} & \multicolumn{2}{c|}{\textit{\textbf{Thread}}} \\ \hline \hline
\textbf{Tool}           & \begin{tabular}[c]{@{}c@{}}\#bin. pat.\end{tabular} & RT   & \#bin                & RT                 & \#bin                & RT                 & \#bin               & RT                  & \#bin                 & RT                   & \#bin                  & RT                   \\ \hline

\textbf{\flowminer}    & 67                                                    & 23s  & 122                  & 36s                & 122                  & 28s                & 430                 & 1.5hr               & 464                   & 134s                 & 460                    & 490s                 \\ \hline

\textbf{\perracotta}   & 30                                                    & 3s   & 2                    & 2.5s               & \multicolumn{2}{c|}{N/A}                  & 7                   & 5min                & xx                    & s                    & 46                     & 62s                  \\ \hline

\textbf{\texada}    & 58                                                    & 2s   & 98                   & 5s                 & \multicolumn{2}{c|}{N/A}                  & 1711                & 280s                & 2052                  & 5s                   & 6601                   & 50s                  \\ \hline


\textbf{\tlmine}   & 12                                                    & 184s & 0                    & 300s               & \multicolumn{2}{c|}{N/A}                  & ssxx                & hrs                 & 26                    & 1528s                &                        &                      \\ \hline


\end{tabular}
\end{table*}
\section{SoC Communication Trace Generation}
\label{sec:tracing}

To the best of our knowledge there is no SoC transaction level communication trace benchmark dataset freely available. Therefore, we describe how a benchmark trace dataset could be produced. We plan to open-source this trace benchmark for scientific community in future. For evaluating the tools under consideration, we divide the trace generation into two major types. The first type is synthetic, that means the traces generated do not necessarily represent a real world SoC communications rather mimics the nature of SoC traces. For example, concurrency of flow in execution, parallelism of event occurrence and recurrence of flow instances are imitated in this trace set. The main motivation of preparing this trace set is that after mining patterns from the tools, we can compare the quality of the mined patterns using the \gt patterns used to generate the traces. The other type of trace we consider is a realistic SoC communication trace captured from simulating a model SoC in  gem5~\cite{gem5_2011} environment. 
This set of traces gives us the opportunity to investigate realistic protocols. The idea is that, we build our confidence on the mining capability of a tool on synthetic trace set and use gem5 traces to see underlying flow specifications of different operations {\tt (write, read, update)} in practice.

\noindent {\bf Synthetic Trace Generation}
This set of traces are produced using ten flows which has $64$ flow instances as \textit{Ground Truth (GT)} patterns. Four such patterns are excerpted in the section~\ref{sec:motivation}. Three set of synthetic traces are generated which are described in Table~\ref{tab:synthetic_traces}. In each set, there are 100 traces, and in each trace, 64 $GT$s are pickup each for a random number times.

\begin{table}[!t]
\renewcommand{\arraystretch}{1.3}
\caption{Synthetic Traces}
\label{tab:synthetic_traces}
\centering
\begin{tabular}{|l||l|}
\hline
\multicolumn{1}{|c||}{\textbf{Trace}} & \multicolumn{1}{c|}{\textbf{Description}}                               
\\ \hline \hline
\emph{SNI}  & \begin{tabular}[c]{@{}l@{}}Single event, non-interleaving pattern traces. A \gt pattern \\ is arbitrarily picked up from the gt pattern pool. It is\\then executed from the start to the end event before the\\next pattern is selected. For example, the flow instance\\patterns from Fig.~\ref{fig:flow-ex} are executed one after another.\end{tabular} 
\\ \hline
\emph{SI}  & \begin{tabular}[c]{@{}l@{}}Single event, interleaved pattern trace, where a few\\patterns are selected (randomly) interleave among\\ themselves to mimic the idea that a single-core SoC is\\running multiple tasks interleaving tasks by priority.\end{tabular}                                  \\ \hline
\emph{MI}  & \begin{tabular}[c]{@{}l@{}}Multi-event, interleaving patterns traces. As the name\\suggests, multiple events might be present in each step,\\such as first step of the trace in (1). This trace set\\tries to mimic a multi-core system running multiple\\tasks in parallel.\end{tabular}                                              \\ \hline
\end{tabular}
\end{table}

\begin{figure}
    \centering
    \includegraphics[width=3in]{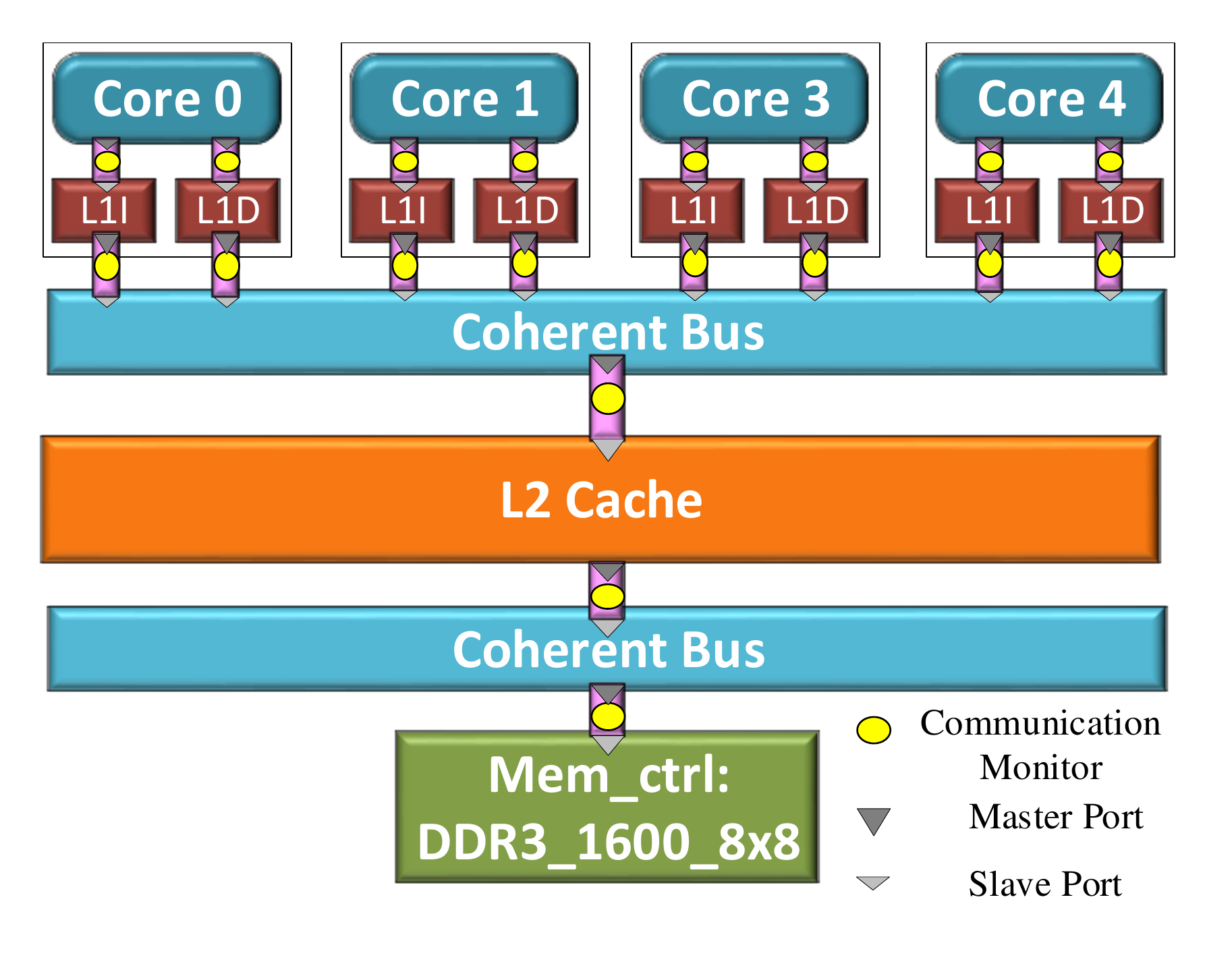}
    \caption{GEM5 model for tracing}
    \label{fig:gem5model}
\end{figure}

\noindent {\bf GEM5 Trace Generation} gem5 is a computer system simulation tool, that simulates in two modes: Full System (FS) simulation and Syscall Emulation (SE). We collect traces from both modes. Fig.~\ref{fig:gem5model} is a simplified model SoC which has four x86 cores. There are private caches: data cache (64kB) and instruction cache (16kB) for each of the cores, and a shared 256kB level 2 cache. There is also a DDR3\_1600\_8x8 memory controller which has a range of upto 4GB of addresses. All these IPs are connected via high speed and concurrent buses that can handle multiple requests from the master components simultaneously. We use 19 communication monitors to observe the \emph{packet}s (unit of communication) in different communication link. Table~\ref{tab:gem5_traces} describes the traces with additional details.

\begin{table}[!t]
\renewcommand{\arraystretch}{1.3}
\caption{Gem5 Traces}
\label{tab:gem5_traces}
\centering
\begin{tabular}{|l||l|}
\hline
\multicolumn{1}{|c||}{\textbf{Trace}} & \multicolumn{1}{c|}{\textbf{Description}}   
\\ \hline \hline
\emph{FS}   & \begin{tabular}[c]{@{}l@{}}One FS mode trace. Boots Linux kernel 4.1.3. No\\workload is executed only kernel boots and exits.\\It has 134 different messages that comprise a length\\of 1.7M steps.\end{tabular}                                  \\ \hline
\emph{Snoop}   & \begin{tabular}[c]{@{}l@{}}Activates simple snoop protocol implemented for\\the traditional memory system. Two SE mode traces \\of 166999 and 189993 steps are captured\\ consisting of 84 messages.\end{tabular}                              \\ \hline
\emph{Threads}  & \begin{tabular}[c]{@{}l@{}}Two SE mode traces, one captured running Paterson\\ algorithm which is a traditional mutex algorithm.\\Another trace is captured for distributing a matrix\\addition. There 129 messages in the traces which in \\combined is 12341234 steps long.\end{tabular}     \\ \hline
\end{tabular}
\end{table}

\section{Observation Summary}
\label{sec:discussion}

One major goal of this study is to find what tools can produce meaning full results for the bench mark traces within a reasonable time limit. All the experiments are conducted on system of 3.2GHz Intel$^{\tiny{\textregistered}}$ core i5 processor, 8GB of RAM. Traces \emph{\textit{FS, Snoop, Thread}} had multiple events in each step. However, these traces are simplified to single event each step format so that tools \perracotta, \texada, \tlmine, \prefixspan and \ylstm can be applied to them. Only tools that were able to complete the mining within 2hrs time limit are listed in Table~\ref{tab:mining_summary}. It lists some experimental findings in terms of number of two event property found and run time. \flowminer and \perracotta do better in terms of finding patterns in most of the cases. Tool \prefixspan runs out of the memory. The main reason behind this is the projected database it creates for every new pattern it mines. The approach \ylstm needs training before any pattern can be extracted. Therefore it is not listed in the summary table. The tool \textit{Trace2model} produces complete execution scenarios that takes more than 3 hours for the two set of synthetic traces, and snooping traces.
\section{Conclusion}
\label{sec:conclusion}

This paper considers some respresentative works from different domains that deals with different types of traces and apply them on SoC communication traces. Comparative study results has been listed. A short review of different tools have been provided. A benchmark of SoC communication trace dataset has also been presented to facilitate more research on this topic. Performance of different tools on this benchmark has been discussed. Crucial advancement SoC validation methods can be achieved with more works on traces incorporating machine learning and other traditional approaches.







\begin{thebibliography}{10}

\bibitem{transaction_based_07}
K.~{Goossens}, B.~{Vermeulen}, R.~v.~{Steeden}, and M.~{Bennebroek}.
\newblock Transaction-based communication-centric debug.
\newblock In {\em First International Symposium on Networks-on-Chip (NOCS'07)},
  pages 95--106, 2007.

\bibitem{ahmed2020mining}
Md~Rubel Ahmed, Hao Zheng, Parijat Mukherjee, Mahesh~C. Ketkar, and Jin Yang.
\newblock Mining message flows from system-on-chip execution traces,
  arXiv-2020.

\bibitem{natasa2020}
Natasha Jeppu, Tom Melham, Daniel Kroening, and John O'Leary.
\newblock Learning concise models from long execution traces.
\newblock In {\em DAC '20}, June 2020.

\bibitem{zheng2021model}
Hao Zheng, Md~Rubel Ahmed, Parijat Mukherjee, Mahesh~C. Ketkar, and Jin Yang.
\newblock Model synthesis for communication traces of system-on-chip designs,
  arXiv-2021.

\bibitem{mining_specs_ammons2002}
Glenn Ammons, Rastislav Bod\'{\i}k, and James~R. Larus.
\newblock Mining specifications.
\newblock POPL '02, page 4–16, New York, NY, USA, 2002. Association for
  Computing Machinery.

\bibitem{Beschastnikh:2011:LEI:2025113.2025151}
Ivan Beschastnikh, Yuriy Brun, Sigurd Schneider, Michael Sloan, and Michael~D.
  Ernst.
\newblock Leveraging existing instrumentation to automatically infer
  invariant-constrained models.
\newblock In {\em Proceedings of the 19th ACM SIGSOFT Symposium and the 13th
  European Conference on Foundations of Software Engineering}, ESEC/FSE '11,
  pages 267--277, 2011.

\bibitem{texada_main}
C.~{Lemieux}, D.~{Park}, and I.~{Beschastnikh}.
\newblock General ltl specification mining (t).
\newblock In {\em 2015 30th IEEE/ACM International Conference on Automated
  Software Engineering (ASE)}, pages 81--92, 2015.

\bibitem{book:ltl_book}
E.~A. Emerson.
\newblock {\em Handbook of Theoretical Computer Science: Volume B: Formal
  Models and Semantics}.
\newblock Elsevier, 995-1072.

\bibitem{wencho_dac2010}
W.~{Li}, A.~{Forin}, and S.~A. {Seshia}.
\newblock Scalable specification mining for verification and diagnosis.
\newblock In {\em Design Automation Conference}, pages 755--760, 2010.

\bibitem{perracotta2006}
Jinlin Yang, David Evans, Deepali Bhardwaj, Thirumalesh Bhat, and Manuvir Das.
\newblock Perracotta: Mining temporal api rules from imperfect traces.
\newblock In {\em Proceedings of the 28th International Conference on Software
  Engineering}, ICSE '06, page 282–291, New York, NY, USA, 2006. Association
  for Computing Machinery.

\bibitem{auto_asser_aspdac2010}
P.~{Chang} and L.~. {Wang}.
\newblock Automatic assertion extraction via sequential data mining of
  simulation traces.
\newblock In {\em 2010 15th Asia and South Pacific Design Automation Conference
  (ASP-DAC)}, pages 607--612, 2010.

\bibitem{Liu:2013}
Lingyi Liu and Shobha Vasudevan.
\newblock Automatic generation of system level assertions from transaction
  level models.
\newblock {\em Journal of Electronic Testing}, 29(5):669--684, Oct 2013.

\bibitem{FournierViger2017ASO}
Philippe Fournier-Viger, J.~Lin, R.~Kiran, Y.~Koh, and R.~Thomas.
\newblock A survey of sequential pattern mining.
\newblock 2017.

\bibitem{agrawal1994}
Rakesh Agrawal and Ramakrishnan Srikant.
\newblock Fast algorithms for mining association rules in large databases.
\newblock In {\em Proceedings of the 20th International Conference on Very
  Large Data Bases}, VLDB '94, page 487–499, San Francisco, CA, USA, 1994.
  Morgan Kaufmann Publishers Inc.

\bibitem{prefixSpan2004}
{Jian Pei}, {Jiawei Han}, B.~{Mortazavi-Asl}, {Jianyong Wang}, H.~{Pinto},
  {Qiming Chen}, U.~{Dayal}, and {Mei-Chun Hsu}.
\newblock Mining sequential patterns by pattern-growth: the prefixspan
  approach.
\newblock {\em IEEE Transactions on Knowledge and Data Engineering},
  16(11):1424--1440, 2004.

\bibitem{bayspec_dac2019}
A.~{Mrowca}, M.~{Nocker}, S.~{Steinhorst}, and S.~{Günnemann}.
\newblock Learning temporal specifications from imperfect traces using bayesian
  inference.
\newblock In {\em 2019 56th ACM/IEEE Design Automation Conference (DAC)}, pages
  1--6, 2019.

\bibitem{yuting_LSTM}
Y.~{Cao}, P.~{Mukherjee}, M.~{Ketkar}, J.~{Yang}, and H.~{Zheng}.
\newblock Mining message flows using recurrent neural networks for
  system-on-chip designs.
\newblock In {\em 2020 21st International Symposium on Quality Electronic
  Design (ISQED)}, pages 389--394, 2020.

\bibitem{deep_spec_2017}
Tien-Duy~B. Le and David Lo.
\newblock Deep specification mining.
\newblock In {\em Proceedings of the 27th ACM SIGSOFT International Symposium
  on Software Testing and Analysis}, ISSTA 2018, pages 106--117, New York, NY,
  USA, 2018. ACM.

\bibitem{spmf2014}
Philippe Fournier-Viger, Antonio Gomariz, Ted Gueniche, Azadeh Soltani,
  Cheng-Wei Wu, and Vincent~S. Tseng.
\newblock Spmf: A java open-source pattern mining library.
\newblock {\em J. Mach. Learn. Res.}, 15(1):3389–3393, January 2014.

\bibitem{gem5_2011}
Nathan Binkert, Bradford Beckmann, Gabriel Black, Steven~K. Reinhardt, Ali
  Saidi, Arkaprava Basu, Joel Hestness, Derek~R. Hower, Tushar Krishna, Somayeh
  Sardashti, Rathijit Sen, Korey Sewell, Muhammad Shoaib, Nilay Vaish, Mark~D.
  Hill, and David~A. Wood.
\newblock The gem5 simulator.
\newblock {\em SIGARCH Comput. Archit. News}, 39(2):1–7, August 2011.

\end{thebibliography}

\end{document}